\documentclass[12pt,preprint]{aastex}
\def\edcomment#1{\iffalse\marginpar{\raggedright\sl#1\/}\else\relax\fi}
\reversemarginpar

\begin{document}
\title{Different Velocity Dependences of Physical Conditions of High- and Low-Ionization Lines in Broad-Line Regions}
\author{Stephanie A. Snedden}
\affil{Apache Point Observatory / New Mexico State University,
2001 Apache Point Rd., Sunspot, NM 88349}
\author{C. Martin Gaskell}
\affil{Dept. Physics and Astronomy, University of Nebraska,
Lincoln, NE 68588}

\begin{abstract}

We present results from a study of high- and low-ionization
emission line ratios as a function of projected velocity for a
sample of eight active galactic nuclei (AGNs). Our results are
based on analysis of high signal-to-noise optical and \emph{Hubble
Space Telescope (HST)} UV spectra. Comparing the emission line
ratios to those predicted by photoionization models indicates that
the physical conditions responsible for the high-ionization
emission lines are consistent with a wind, whereas those of the
low-ionization lines are consistent with a virialized disk.

\end{abstract}

\section{Introduction}

It is evident from reverberation mapping studies that high- and
low-ionization emission lines in AGNs show different time-lag
responses to continuum variability. Presumably, they are generated
at different distances from the central engine (Korista et al;
1995, Santos-Lle\'{o} et al.; 1997, among others). However, it is
still a matter of debate whether the kinematics of the high- and
low-ionization regions are fundamentally different; e.g., is the
high-ionization gas associated with a wind and the low-ionization
gas with a disk? In this study we look for evidence of virialized
motion from the emission lines of both high- and low-ionization
gas of AGNs. We do this by looking at line ratios as a function of
projected velocity across the emission line profiles.

\section{Data and Analysis}

For this study, we measured emission line ratios as a function of
projected velocity for eight AGNs with high signal-to-noise
optical and UV spectra. The sample is listed in Table 1. The
optical spectra, provided by Giovanna Stirpe, include the
H$\alpha$ and H$\beta$ lines. [S\,II] $\lambda \lambda 6716,
6731$, [N\,II] $\lambda 6583$, [O\,I] $\lambda 6364$ + [Fe\,X]
$\lambda 6374$, [O\,III] $\lambda \lambda 4959, 5007$ were
deblended from the Balmer lines, the narrow-line components were
removed, and a power law continuum was subtracted, all as
described in Stirpe, 1991. The $\emph{HST}$ $\emph{FOS}$ UV
spectra were deblended of N\,V $\lambda 1240$, Al\,III] $\lambda
1857$, Si\,III] $\lambda 1892$, and a local power law continuum
was subtracted from the Ly$\alpha$ to C\,III] region. Next, we
integrated all emission line fluxes over a high-velocity component
centered at $\pm 4000 km\cdot s^{-1}$, an intermediate-velocity
component centered at $\pm 2000 km\cdot s^{-1}$ and a low-velocity
component centered at $0 km\cdot s^{-1}$. The ratios
Ly$\alpha$/H$\alpha$, H$\beta$/H$\alpha$, Ly$\alpha$/C\,IV and
C\,III]/C\,IV for each velocity component were then calculated. We
used $\emph{CLOUDY}$ v90.04 (Ferland et al.; 1998) to generate
grids of emission line ratios as a function of hydrogen density,
$n_H$, and surface photon flux capable of ionizing hydrogen,
$\it{\Phi_H}$. The measured emission line ratios were compared to
the predicted values to determine $n_H$ and $\it{\Phi_H}$ as a
function of velocity for the objects in our sample.

Figures 1 and 2 show $n_H$ and $\it{\Phi_H}$ averaged over the
objects in our sample, normalized to the value for the low
velocity gas. Figure 1 shows the predicted physical conditions as
a function of velocity for the low-ionization gas, as deduced from
the hydrogen lines. Note that the normalized surface photon flux
is plotted to the one-fourth power. If the photon flux ratio
 is linear with respect to the absolute value of the velocity in this plot, it implies the line-emitting gas obeys the inverse-square law.
 Figure 2 shows the same for the high-ionization gas, as deduced
from the carbon lines and Ly$\alpha$. However, for this case, the
normalized surface photon flux is $\it{not}$ raised to the
one-fourth power, since $\it{\Phi_H}$ is constant with respect to
velocity for the high-ionization gas. In all figures the errorbars
show cumulative errors, including scatter in the average ratios
across the sample as well as uncertainty in the physical
conditions as deduced by matching observed line ratios to the
photoionization models. The latter error results from
uncertainties in deblending and continuum subtraction.

\begin{table}

\begin{center}
\begin{tabular}{l}
\hline \\
Object \\
 \hline \\
 B2 2201+31A \\
 Fairall 9 \\
 NGC 3783 \\
 NGC 5548 \\
 PG 1116+215 \\
 PG 1211+143 \\
 PG 1351+640 \\
 Pks 2251+113 \\
\hline
\end{tabular}

\end{center}
\caption{The sample.}
\end{table}

\begin{figure}
\plottwo{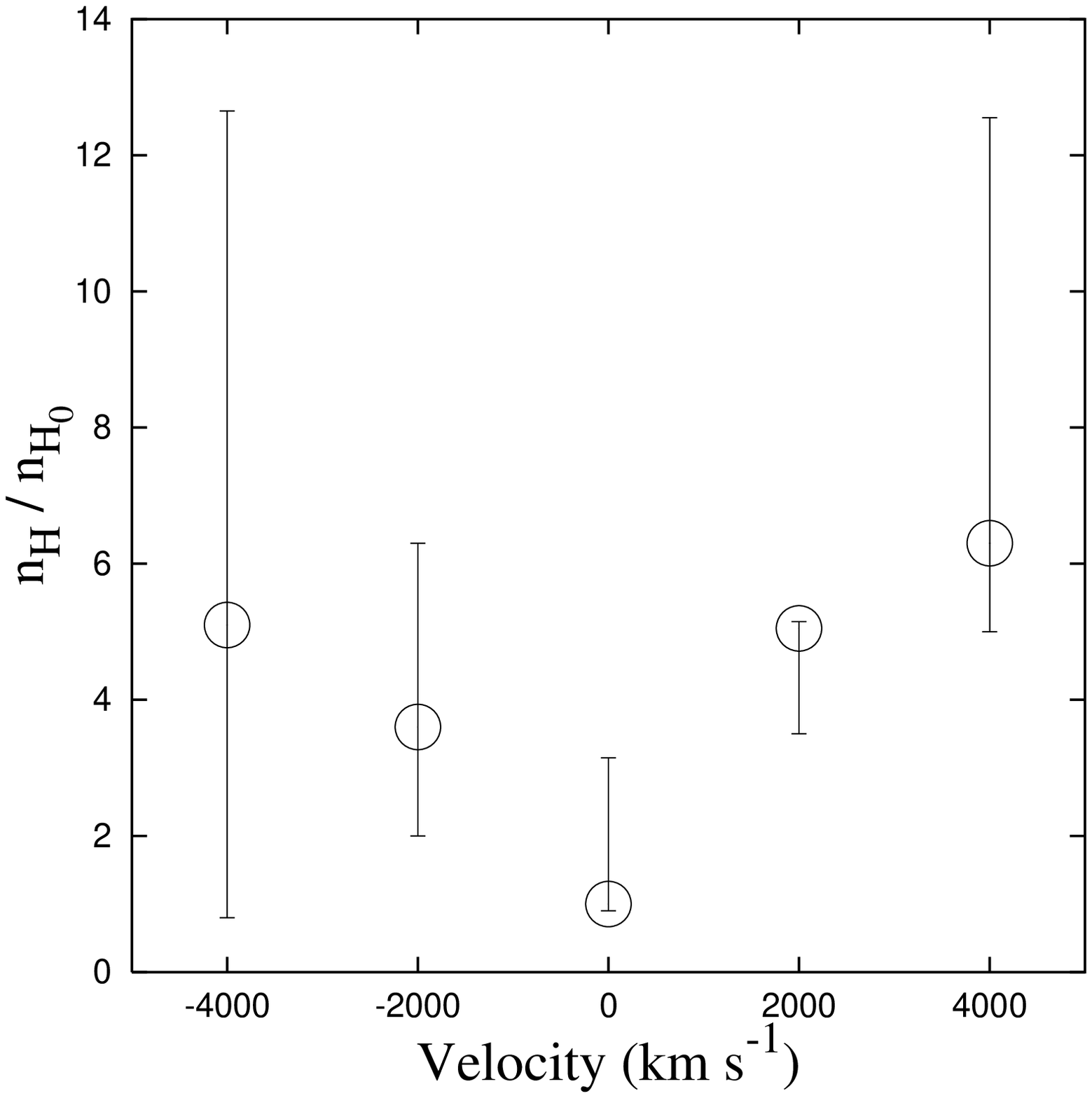}{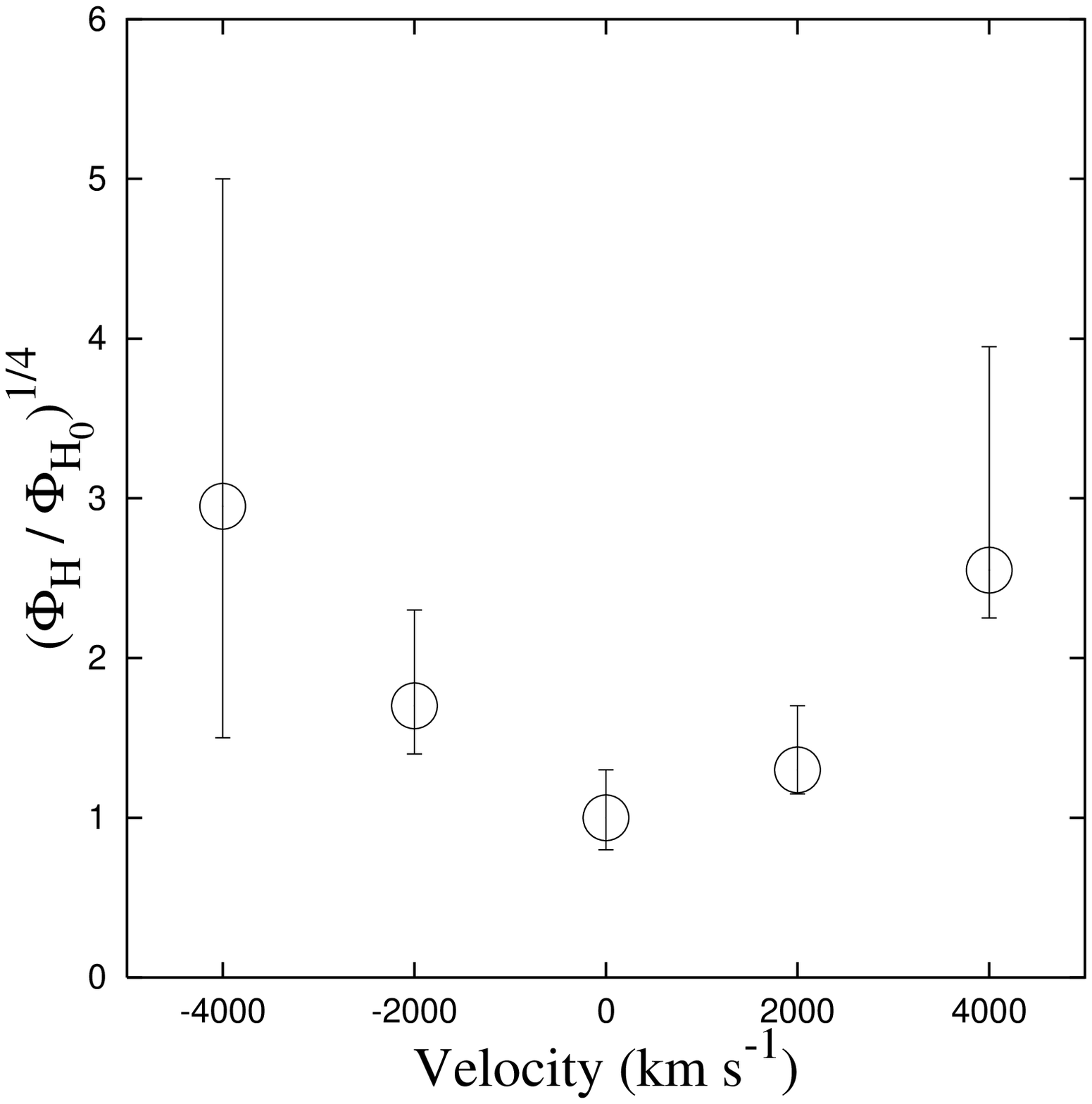} \caption{$a)$ Average
density, $n_H$, for the entire sample, deduced from hydrogen lines
and normalized to the low velocity value, $n_{H_0}$, as a function
of velocity. $b)$ Average surface photon flux, $\it{\Phi_H}$, for
the entire sample, deduced from hydrogen lines and normalized to
the low velocity value, $\it{\Phi_0}$, as a function of velocity.
We plot this ratio to the one-fourth power.}
\end{figure}

\begin{figure}
\plottwo{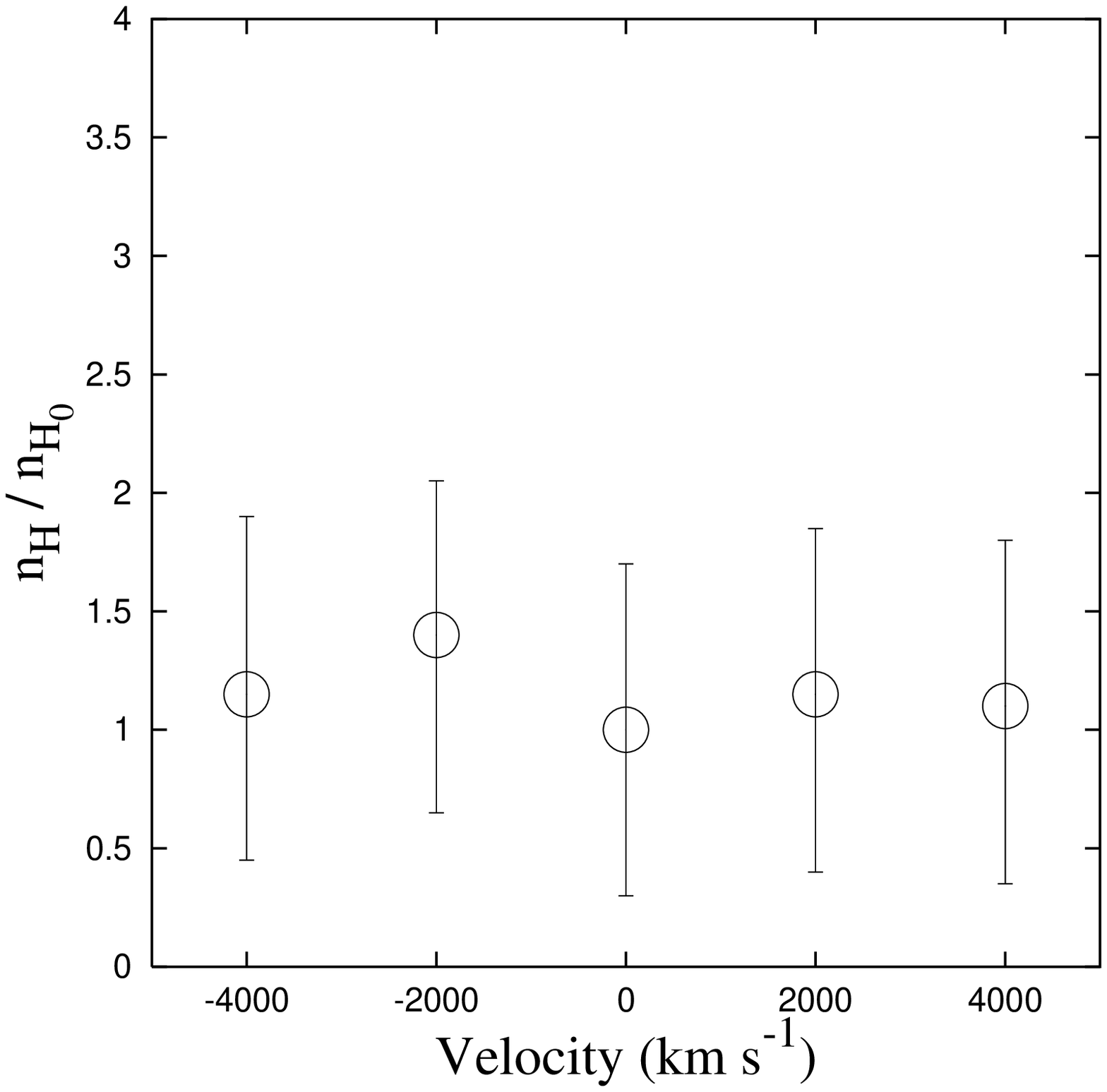}{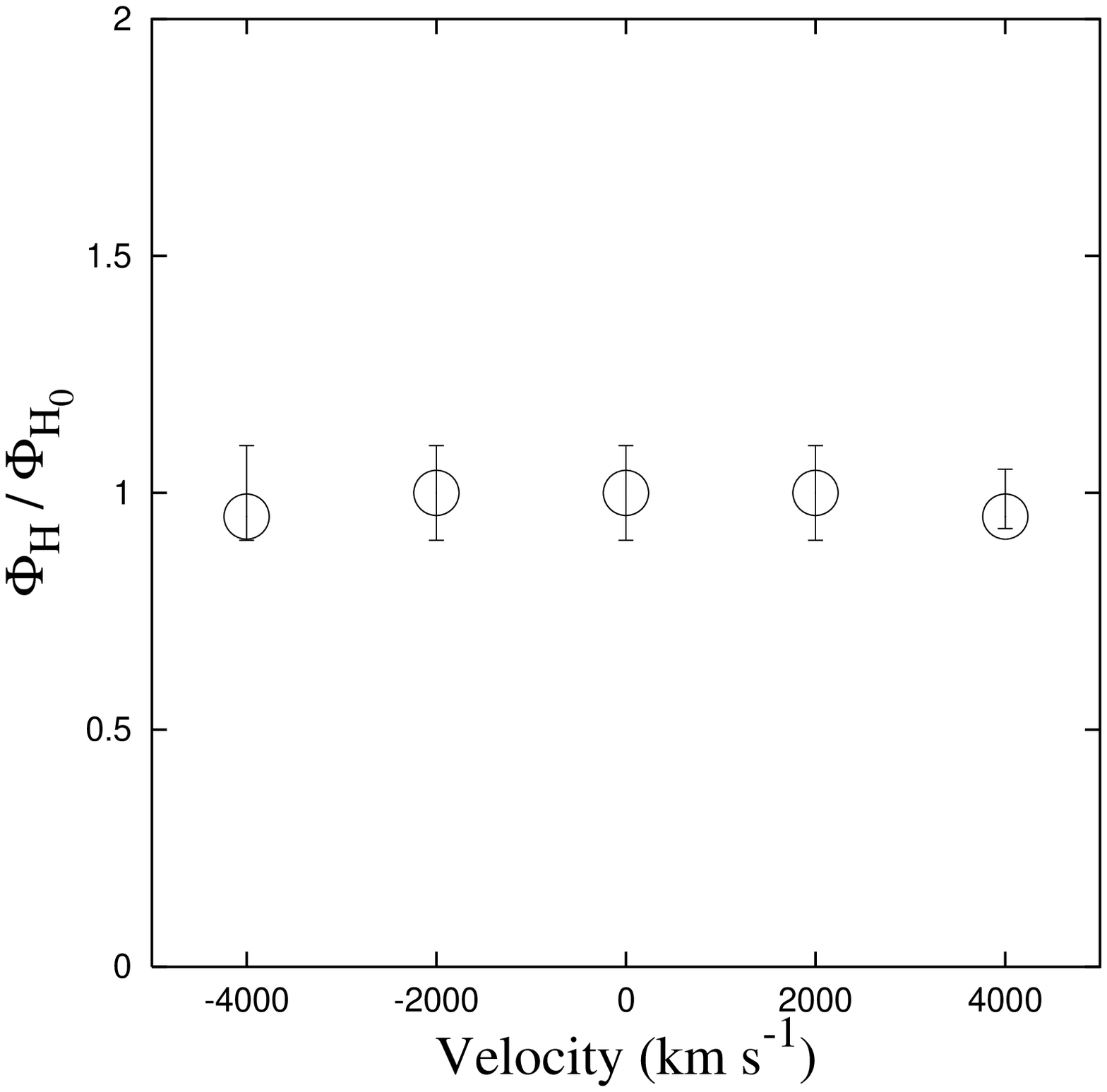}

\caption{ $a)$ Average density, $n_H$, for the entire sample,
deduced from high-ionization lines and normalized to the low
velocity value, $n_{H_0}$, as a function of velocity. $b)$ Average
surface photon flux, $\it{\Phi_H}$, for the entire sample, deduced
from high-ionization lines and normalized to the low velocity
value, $\it{\Phi_0}$, as a function of velocity.}
\end{figure}

\section{Discussion}

A comparison of Figures 1 and 2 shows a fundamental difference in
the photoionization model predictions of physical conditions of
the low- and high-ionization gas. The low-ionization
Balmer-line-emitting gas (Figure 1), shows a clear change in both
$n_H$ and $\it{\Phi_H}$ as a function of velocity. The higher
velocity gas is denser than low velocity gas. There is also a
trend toward higher $\it{\Phi_H}$ as projected velocity increases,
and the trend fits the inverse-square law within the errors. The
high-ionization gas, however, shows no such trend (Figure 2). Both
$n_H$ and $\it{\Phi_H}$ are constant as a function of projected
velocity.

\section{Conclusions}

The results shown in Figure 1 are consistent with the Balmer line
emitting gas being virialized. The lower velocity low-ionization
gas sees a smaller $\it{\Phi_H}$ in accord with the inverse-square
law. Figure 2, however, is evidence of fundamentally different
kinematics for the high-ionization gas that produces C\,IV. In
this case, the independence of physical conditions with respect to
projected velocity, averaged over the sample, is consistent with
gas motion that is not virialized.

These results are consistent with the high-ionization lines
arising in a wind and the low-ionization lines arising in a disk.

\end{document}